\documentclass[prb,aps,preprint,showpacs]{revtex4}
\usepackage{epsfig}
\usepackage{amssymb}
\usepackage{amsmath}
\usepackage{graphicx}
\usepackage{amsfonts}
\usepackage{color}
\usepackage{textcomp}

\usepackage{epstopdf}
\usepackage{pifont}

\draft
\begin{document}

\title{Interaction Effects in Conductivity of a
Two-Valley Electron System in High-Mobility Si Inversion Layers}
\author{N.~N.~Klimov$^{1,3}$, D.~A.~Knyazev$^2$,
O.~E.~Omel'yanovskii$^2$, V.~M.~Pudalov$^2$,
H.~Kojima$^1$, and M.~E.~Gershenson$^1$}
\address{
$^1$Department of Physics and Astronomy, Rutgers University, New
Jersey 08854, USA.\\
$^2$P.~N.~Lebedev Physical Institute,  Moscow 119991, Russia.\\
$^3$ P.~N.~Lebedev Physics Research Center, Moscow 119991, Russia. \\
}

\begin{abstract}
We have measured the conductivity of high-mobility (001)~Si
metal-oxide-semiconductor field effect transistors (MOSFETs) over
wide ranges of electron densities $n = (1.8-15)\times
10^{11}$\,cm$^{-2}$, temperatures $T = 30$\,mK -- 4.2\,K, and
in-plane magnetic fields $B_\parallel= (0-5)$\,T.  The experimental
data have been analyzed using the theory of interaction effects in
the conductivity $\sigma$ of disordered 2D systems. The parameters
essential for comparison with the theory, such as the intervalley
scattering time and valley splitting, have been measured  or
evaluated in independent experiments. The observed behavior of
$\sigma$, including its quasi-linear increase with decreasing $T$
down to $\sim 0.4$\,K and its downturn at lower temperatures, is
in agreement with the theory. The values of the Fermi-liquid
parameter obtained from the comparison agree with the
corresponding values extracted from the analysis of Shubnikov-de
Haas oscillations
based on the
theory of magnetooscillations in interacting 2D systems.
\end{abstract}

\pacs{PACS: }

\date{\today}

\maketitle

\section{INTRODUCTION}
The beginning of the 80's witnessed a triumph of the one-parameter
scaling theory of localization \cite{AALR79,wegner79} and the
theory of electron-electron 
 interactions in disordered
conductors (for a review, see Ref.~\onlinecite{alshuler80}).  As a
result, the peculiar low-temperature behavior of the conductivity
of numerous low-dimensional systems has been successfully
attributed
to 
 quantum interference effects (see, e.g.,
Ref.~\onlinecite{altshuler_gershenson87}).  Curiously, application
of these ideas to the two-dimensional (2D) electron liquid in
Si~MOSFETs - one of the most ubiquitous 2D systems - remained a
challenge for more than 25 years.  In the early experiments
\cite{bishop80,dolgopolov84} with low-mobility ($\mu\sim
0.1\,$m$^2$/Vs) structures in the regime of relatively high
electron densities
($n>10^{12}$cm$^{-2}$), a decrease
of the conductivity with cooling has been observed, apparently in
 qualitative agreement with the ideas of
weak localization \cite{AALR79,gorkov79} and electron-electron
interactions \cite{alshuler80}. However, the quantitative
description of this behavior remained a problem \cite{burdis88}.
The disagreement with the theory became qualitative with the
advent of high-mobility ($\mu\geq 2\,$m$^2$/Vs)
structures:  
  the
 low$-T$
conductivity of high-$\mu$  Si~MOSFETs increased with decreasing
temperature \cite{cham&wheeler80}, in striking contrast to the
behavior of many other 2D systems.  This ``metallic'' behavior of
the conductivity is especially pronounced at low electron
densities $n \sim (1-2)\times 10^{11}$cm$^{-2}$, where a
 five-fold increase of $\sigma(T)$ with cooling was observed
\cite{smith86,zavaritskaya87,virodov88,krav94}. Later, the
quasi-linear increase of $\sigma(T)$  was observed in
many high-mobility 2D systems in the ``dilute'' regime, including
GaAs heterostructures (both $p$-type \cite{14,15,16,17} and
$n$-type \cite{18,19}), Si/SiGe \cite{20}, and AlAs \cite{21}
quantum wells.

The subsequent development of theory and experiment led to
 significant progress in our understanding of the low-temperature
transport in high-mobility systems in the regime of low electron
densities.  In the mid-80's, the quasi-linear ``metallic''
dependence observed in the ballistic interaction regime  $T\tau\gg
1$ ($\tau$ is the transport mean free time, here and below we set
 $\hbar=k_B=e=1$) was attributed \cite{22} to
weakening of the screening of the
scattering potential with increasing $T$. Observation of a strong
negative magnetoconductance induced by the
in-plane magnetic field \cite{23} and renormalization of the
effective electron mass and $g-$factor in these structures
\cite{24,gm02} also suggested that the electron-electron
interactions play an important
role in this phenomenon.  More recently, Zala, Narozhny, and
Aleiner (ZNA) \cite{26} developed a theory that took into account
all interaction contributions to the conductivity, including the
exchange ones. This theory offers a unified approach to both
ballistic ($T\tau\gg 1$) and diffusive ($T\tau \ll 1$) interaction
regimes by considering the quantum interference between electron
waves scattered off a short-range random potential ``dressed'' by
Friedel oscillations of the electron density. The theory was
extended for the case of a long-range scattering potential by
Gornyi and Mirlin (GM) \cite{27}.  The theories \cite{26,27}
naturally incorporate the Altshuler-Aronov 
 results for the
interaction corrections to the conductivity in the diffusive
regime \cite{alshuler80}.

For the
diffusive
regime,
a more general
approach to interacting systems based on
 the nonlinear $\sigma$-model
has been developed by Finkel'stein \cite{fink}.
Recently, the renormalization
group (RG) equations of this theory \cite{fink02,pf_science}
(obtained in the first order in $\frac{1}{\pi\sigma}$ and in all orders in
interaction) have been
compared with
the conductivity of Si~MOSFETs at low electron
densities
\cite{fink02,knyazev06,krav_nature_07,critical}.

The RG equations \cite{fink,castellani84,CdCL98,fink02,pf_science}
describe the length scale (temperature) evolution of the resistivity
and interaction parameters
for a 2D electron system in the diffusive regime
\cite{fink}.
However, at high electron densities,
the temperature range corresponding to the diffusive regime
 shrinks. In contrast, the theory of
interaction corrections \cite{26,27} is applicable over a wider $T$
range (that includes both ballistic and diffusive regimes) provided
$\sigma \gg 1$ and $\Delta\sigma/\sigma
\ll 1$; these assumptions are well justified
at high densities.

The theories\cite{26,27} predict that the magnitude and sign of the the interaction correction $\Delta\sigma(T,B)$
  is determined by the value of the
Fermi-liquid parameter $F_0^\sigma$ (which can be found by
measuring the Shubnikov-de Haas (SdH) oscillations in weak
magnetic fields perpendicular to the plane of a 2D structure
\cite{24,gm02} or the magnetoresistance in strong in-plane
magnetic fields\cite{28}). In particular, it is expected that the
$\sigma(T)$ dependence becomes ``metallic'' when $F_0^\sigma$  is
negative and its absolute value exceeds a certain threshold (see
Sec. II).

The experimental studies of the conductivity in various
low-carrier-density 2D systems in the ballistic (high-temperature)
regime are in agreement with the ZNA and GM theories.  The
``metallicity'' in all these systems is enhanced at low $n$ due to
an increase of the absolute value of  $F_0^\sigma$  (see, e.g.,
Refs.~\onlinecite{29,30,31}). In Si~MOSFETs, the interaction
effects are especially strong due to the presence of two nearly
degenerate valleys in the electron spectrum \cite{fink02}.
This enhancement, however, diminishes if the
temperature $T$ becomes smaller than the  valley splitting
$\Delta_V$ and
intervalley scattering rate $\tau_{v}^{-1}$.
 As a result, with lowering $T$, the ``metallic''
 dependence of $\sigma(T)$ is expected to  become weaker or even
 to
 be replaced with an
``insulating'' one. To the best of our knowledge, this behavior
has not been observed for Si~MOSFETs prior to our work.

This paper aims to study in detail the conductivity of 2D electron liquid in
high-mobility (001)~Si~MOSFETs over a wide temperature range ($T = 0.03-4.2$\,K)
that includes both the {\em diffusive} and {\em ballistic} regimes.
In particular, we observed for the first time that the ``metallic'' increase of
$\sigma$ with cooling is followed by the downturn of  $\sigma(T)$ at lower temperatures.
 For the purpose of comparison with the ZNA theory \cite{26}, we studied the
 range of not-too-low densities,  $n=(1.8-15)\times 10^{11}$\,cm$^2$,
 where the temperature and magnetic
 field dependences   $\Delta\sigma(T,B)$ can still be treated as small corrections
 to the Drude conductivity  $\sigma_D$.  In principle, no fitting parameters are
 required for comparison with the theory, because we have measured
 $F_0^\sigma$,  $\Delta_V$, and $\tau_{v}$
 in independent experiments.  However, below we take a slightly different approach:
 we obtain the $F_0^\sigma(n)$  values from fitting the $\Delta\sigma(T,B_\parallel)$
 dependences   with the ZNA theory \cite{26}, and
 show that these values are consistent with the corresponding values extracted
 from the analysis of SdH oscillations \cite{gm02}.
 We have also revealed  shortcomings of earlier
analysis of $\Delta\sigma(T)$, 
 reanalyzed the available data, and  compared the extracted values of
 $F_0^\sigma(n)$  with corresponding values from other measurements.
 We conclude that the experimental
 data are well described by the  theory
of interaction corrections
 \cite{26}
  at
 intermediate temperatures $T\approx (0.3-4.2)$\,K.
 For a quantitative analysis at ultra-low temperatures ($T \leq 0.3$\,K),
the interaction correction theory
should be modified by taking into account finite
intervalley scattering rates.

The paper is organized as follows.  In Section II we briefly summarize
the theoretical results \cite{26} for the interaction corrections to the
conductivity of a two-valley system.  The experimental data are  
 presented in Section III, along with the
data analysis and discussion.  The summary is 
 given in Section IV.

\section{INTERACTION CORRECTIONS TO THE CONDUCTIVITY}
\subsection{Temperature dependence of the conductivity in zero magnetic field}
In the ZNA theory \cite{26}, the corrections   $\Delta\sigma_{ee}$
to the Drude conductivity 
$\sigma_D=n\tau/m_b$ ($m_b\approx 0.205 m_e$ is the electron
band mass in (001)~Si~MOSFETs;
for more detail, see discussion in section \ref{m_b} and
references therein) were calculated in both ballistic 
 and diffusive 
  regimes 
for all orders of the
interaction strength and 
the leading order in
$1/(E_F \tau)$ and  $T/E_F$.  In particular, the
theory reproduces the Altshuler-Aronov
correction\cite{alshuler80} to the conductivity in
the diffusive regime.  To adapt the theoretical results \cite{26}
to the case of (001)~Si~MOSFETs, one should take into account that
the electron spectrum in this system has two almost degenerate
valleys \cite{ando}. In zero magnetic field,
$\Delta\sigma_{ee}(T)$ for a system with two degenerate valleys in
the absence of intervalley scattering can be written as follows
\cite{30}:
\begin{equation}
\Delta\sigma_{ee}(T)= \delta\sigma_C(T) +15 \delta\sigma_T(T).
\label{1}
\end{equation}
Here $\delta\sigma_C$ is the so-called ``charge''
contribution which combines Fock correction
and the singlet part of Hartree correction, and  $\delta\sigma_T$
is the ``triplet'' contribution
due to the triplet part of Hartree term.  The valley index can be considered as a
pseudo-spin in multi-valley systems \cite{fink02}, and the
valley degeneracy determines the
number of triplet terms due to the spin exchange processes between electrons in
different valleys.  For a system with two degenerate valleys, the total number
of interaction channels is  $4\times 4=16$, among them 1 singlet and 15 triplet terms
(for comparison, there are 1 singlet and 3 triplet terms for a single-valley system).

Below we assume that the scattering potential is short-ranged which is
relevant to Si MOSFETs. According to Ref.~\onlinecite{26}, the
charge term does not depend on the details of interactions:
\begin{equation}
\delta\sigma_C=\frac{1}{\pi}\left \{ (T\tau ) \left ( 1-\frac{3}{8}f(T\tau) \right ) -
\frac{1}{2\pi} \ln \frac{E_F}{T} \right \},
\label{2}
\end{equation}
whereas the magnitude and sign of the triplet term is controlled by
the Fermi-liquid
 parameter \cite{26} $F_0^\sigma$:
\begin{equation}
\delta\sigma_T=\frac{1}{\pi}\left \{ (T\tau )\frac{F_0^\sigma}{1+F_0^\sigma}\left
(1-\frac{3}{8}t(T\tau; F_0^\sigma) \right ) - \left (1-\frac{\ln \left( 1+F_0^\sigma \right)}{F_0^\sigma}
  \right )
\frac{1}{2\pi} \ln \frac{E_F}{T} \right \}.
\label{3}
\end{equation}
The functions $f$ and $t$
in Eqs.~(\ref{2},\ref{3}) describe the crossover between the diffusive
($\Delta\sigma_{ee} \propto \ln T$) and ballistic
($\Delta\sigma_{ee} \propto T$ ) regimes; outside the crossover
region, they change the value of $\Delta\sigma_{ee}$  by only a
few percent. The explicit expressions for these functions can be
found in Ref.~\onlinecite{26}. The diffusive-ballistic crossover
is expected over some temperature range near
\begin{equation}
T^*=\frac{1+F_0^\sigma}{2\pi \tau}.
\label{4}
\end{equation}
Equations (\ref{2},\ref{3}) describe the quantum corrections
in a system with the conductance
$\sigma \gg 1/2\pi$
at temperatures well below the Fermi
 energy $\left [T\ll \left(1+F_0^\sigma\right)^2 E_F \right ]$.

The sign and the magnitude of $\Delta\sigma_{ee}$
is controlled by the Fermi-liquid 
 parameter $F_0^\sigma$.
For a rough estimate, deeply in the ballistic regime
the $\ln T$ terms and the crossover functions $t$ and $f$ in Eqs.~(\ref{2},\ref{3})
can be omitted.
For example,   at $T\tau=10$,
the functions $f$ and $t$  contribute $\approx
4\%$ and $\approx 11\%$, respectively, to the linear-in-$T\tau$
ballistic terms of Eqs.~(\ref{2},\ref{3}).
By neglecting these
terms, we find that for a single-valley system
the linear dependence $\Delta\sigma_{ee}(T)$ in the
ballistic regime becomes ``metallic'' ($d\sigma/dT <0$) at
$3F_0^\sigma/(1+F_0^\sigma) <-1$ or $F_0^\sigma <-0.25$, whereas
for a system with two degenerate valleys, the ``metallic''
$\Delta\sigma_{ee}(T)$ dependences are expected for $15
F_0^\sigma/(1+F_0^\sigma) <-1$  or  $F_0^\sigma <-0.06$.
 Thus, the valley degeneracy extends the range of  $F_0^\sigma (n)$,
 and hence, the range of carrier
 densities $n$ where the conduction exhibits  the ``metallic'' behavior.

\subsection{Magnetoconductivity in the in-plane magnetic field}
The in-plane magnetic field, being coupled mostly to electron
spins, provides a useful tool for exploring the interaction
effects in the low-temperature conductivity of Si~MOSFETs
\cite{23}.  When the Zeeman energy $E_Z=g_b\mu_B B$ ($g_b = 2$ is
the bare  $g$-factor, $\mu_B$  is the Bohr magneton) becomes much
greater than $T$, the number of triplet terms that contribute to
$\Delta\sigma_{ee}(T)$  is reduced from 15 to 7. Similar reduction
of triplet terms is expected for a valley splitting  $\Delta_V
>T$. These two effects have been accounted by the theory of
interaction corrections  \cite{26,30}; in the presence of the
magnetic field and/or valley splitting the interaction correction
to the conductivity can be expressed as follows \cite{30}:

\begin{eqnarray}
\Delta\sigma_{ee}(T,\tau,F_0^\sigma, B_{\parallel}, \Delta_V)&=&
\Delta\sigma_{ee}(T)+
2\Delta\sigma^Z(E_Z,T)\nonumber\\
&+&2\Delta\sigma^Z(\Delta_V,T)+
\Delta\sigma^Z(E_Z+\Delta_V,T)+
\Delta\sigma^Z(E_Z-\Delta_V,T), \quad
\label{5}
\end{eqnarray}
 where $\Delta\sigma_{ee}(T)$  is given by Eq.~(\ref{1}). All the terms
$\Delta\sigma^Z(Z,T)$  have a form

\begin{eqnarray}
&&\Delta\sigma^Z(Z,T)\equiv
\sigma(Z,T)-\sigma(0,T)=\delta\sigma_b(Z)+\delta\sigma_d(Z)=\nonumber\\
&&\frac{1}{\pi}\left\{\left[\frac{2F_0^\sigma}
{1+F_0^\sigma}(T\tau)K_b\left(\frac{Z}{2T},F_0^\sigma
\right)\right]+ \left[K_d\left(\frac{Z}{2\pi T},F_0^\sigma \right)
\right]+ m\left(Z\tau,T\tau;F_0^\sigma \right)\right\},
\label{6}
\end{eqnarray}
if the relevant energies $Z\ll E_F$ ($Z$ stands for $E_Z$,
$\Delta_V$, and combinations $E_Z \pm \Delta_V$). The explicit
expressions for the functions $K_b$ and $K_d$  are given in
Ref.~\onlinecite{26}.  In particular, Eq.~(\ref{6})
describes the interaction-driven magnetoconductivity in
 the magnetic fields which
are much weaker than the field of full spin polarization of a system.
Below we will neglect the function $m(0,T\tau; F_0^\sigma$)
which describes the crossover between
the ballistic and diffusive regimes: this function 
is numerically small and does not modify the value of
$\Delta\sigma(Z,T)$ outside the ballistic-diffusive crossover
region by more then one percent.

It is worth mentioning
 that in the framework of the RG theory,
the magnetoconductance can  also be described by the Castellani-Di
Castro-Lee formula \cite{CdCL98,burmi06} which is equivalent to
Eq.~(\ref{5}) in the diffusive limit at $\Delta_V=0$.
However, for the analysis of our magnetoconductivity data measured
over a wide $T$-range that includes both diffusive and ballistic
regimes, the interaction correction theory \cite{26} is more
appropriate than the RG theory \cite{CdCL98}.

The interaction correction theory \cite{26} (as well
as the RG theory \cite{fink02,pf_science}) does not take into
account intervalley scattering. This approximation is valid when
the intervalley scattering rate $\tau_v^{-1}$ is much smaller than
$T$.  In the low-temperature limit  $T\ll \tau_v^{-1}$, the
electron states in different valleys are completely intermixed at
the time scale $\sim T^{-1}$ and the correction
$\Delta\sigma_{ee}(T)$ for a two-valley system  is
expected to coincide with that for a single-valley system.

Since the interaction corrections to the
conductivity $\Delta\sigma_{ee}(T,B)$
 depend 
  on several parameters such as $\tau, F_0^\sigma,
  \Delta_V$ and $\tau_v$, for
 testing the theoretical results, it is crucial to
determine these parameters
 in independent measurements.
 This program is realized in the following section.

\section{SAMPLE CHARACTERIZATION AND DATA ANALYSIS}
\subsection{Sample Characterization}
We have studied the temperature and magnetic field dependences of
the conductivity for high-mobility (001)~Si~MOSFETs, which
demonstrate the ``metallic'' quasi-linear $\sigma(T)$ dependences
at intermediate temperatures over a wide range of electron
densities $n$. In this paper we present the data for two (001)
Si~MOSFETs, Si6-14 and Si1-46, with the gate oxide thickness $190
\pm  20$\,nm and peak mobility $\mu(0.1{\rm K})\approx
2$\,m$^2$/Vs.  The  $\sigma(T,B)$ dependences were measured over
the temperature range $T=0.03-4.2$\,K using a standard
low-frequency four-terminal technique. The measuring current was
chosen sufficiently small ($I\sim 1-3$\,nA) to avoid overheating
of electrons within this temperature range \cite{cooling02}. Our
experimental set-up allowed us to independently control the
magnetic field  normal to the plane of 2D layer ($B_\perp \approx
-1.5 ... +1.5$\,T) and the in-plane magnetic field ($B_\parallel
\approx -8... +8$\,T); this cross-magnetic field technique has
been described in Ref.~\onlinecite{37}. Unless otherwise stated,
an in-plane magnetic field of $B_\parallel  = 0.02$\,T was applied
to quench the superconductivity in the current/voltage contact
pads and the gate electrode which are made of thin aluminum films.

\subsubsection{Effective mass $m^*$ and $g-$factor.
Interaction corrections to the magnetooscillations.}
\label{m*}
 The electron density $n$, the electron temperature $T$, 
 the effective electron mass $m^*$, and $g-$factor $g^*$ have been
found from
measurements of 
SdH oscillations (see also
Ref.\,\onlinecite{gm02}). For fitting the oscillations,
as the first step of the analysis, we have used 
Lifshitz-Kosevich (LK)
formula \cite{LK} which is valid for non-interacting 2D electrons
if the amplitude of oscillations is small \cite{40}:
\begin{eqnarray}
\frac{\delta\sigma_{xx}(\omega_c,T)}{\sigma_D}&=&\sum_s
A_s^{\rm LK}(\omega_c,T)\cos\left[ \pi s\left(\frac{2E_F}{\omega_c}
-1\right)\right] \cos\left[ \pi s \frac{ g^*m^*}{2m_e}\right]
\cos\left[ \pi s \frac{\Delta_V}{\omega_c}\right].
\nonumber\\
A_s^{\rm LK}(\omega_c,T)&=&-4\exp\left(-\frac{2\pi^2 s
T_D}{\omega_c}\right) \frac{2\pi^2 s T/\omega_c}{\sinh(2\pi^2 s
T/\omega_c)}
\label{SdHbyLK}
\end{eqnarray}
Here  $\omega_c=B_\perp/m^*$ is the cyclotron frequency,
$T_D=1/2\pi \tau_D$ is the Dingle temperature, $\tau_D$ is the
elastic quantum scattering time. Figure 1 shows that the
temperature dependence of the amplitude of the first harmonic,
$A_1(T)$, is in agreement with Eq.~(\ref{SdHbyLK}) down to the lowest temperatures;
this indicates that the electrons are not overheated
(with respect to the thermal bath) by
the bias current and/or noise.

The $F_0^\sigma(r_s)$ values can be found from the measurements of
the renormalized $g$-factor: $F_0^\sigma=(g_b/g^*)-1$. Here
$r_s\equiv 1/\overline{a}^*_B \sqrt{\pi n}$  is the dimensionless
ratio of the Wigner-Seitz radius to the effective Bohr radius,
 $\overline{a}_B^* =\overline{\kappa}/m e^2 $,
  $\overline{\kappa}=7.7$ is
 the average dielectric constant of Si and SiO$_2$, and
$m=0.19m_e$ is the 
electron band mass in bulk Si\cite{ando,019me}.
In our experiments \cite{gm02}, $g^*(n)$ was obtained from the
analysis of SdH oscillations as the ratio of two quantities:
the measured renormalized electron spin susceptibility
$\chi^*/\chi_b =g^*m^*/g_b m_b$ and the effective mass $m^*$
($\chi_b$ 
is the bare value of 
spin susceptibility). Observation of the beatings of SdH
oscillations in crossed magnetic fields offers a straightforward
(model-independent) method of finding $g^*m^*$ \quad
\cite{gm02,37}. On the other hand,
an estimate of $m^*$ is based on a model-dependent analysis of
the damping factor for the first harmonic of SdH oscillations,
$A_1(T,B_\perp=const)$.  According to the LK theory, the damping
factor can be expressed as

\begin{equation}
-\ln\left[A_1^{\rm LK}(T,B_\perp) \right] \frac{B_\perp} {2\pi^2 m^*} \approx
\left(T+T_D \right). \label{LK}
\end{equation}
Our experiments show that $\ln A_1(T,B_\perp)$ varies linearly
with $T$ within the experimental range $T = (0.4-0.8)$\,K (see
Fig.~3 in Ref.\,\onlinecite{gm02}); this however does not prove
the applicability of the LK theory,
which disregards the interaction effects.  It was
recently shown \cite{40,41} that due to the interference between
electron-electron and electron-impurity interactions, the damping
factor acquires an additional term
in both the diffusive and ballistic
regimes
\begin{equation}
-\ln\left[A_1(T,B_\perp) \right] \frac{B_\perp} {2\pi^2 m^*} = \left(T+T_D
\right)-\alpha(T),
\label{LK+GM}
\end{equation}
where
\begin{equation}
\alpha(T)=-T\frac{\delta m^*}{m^*}-T_D\left(\frac{\delta m^*}{m^*} -
\frac{\delta \tau_D^*}{\tau_D^*}\right),
\label{alpha}
\end{equation}
and
\begin{eqnarray}
\frac{\delta m^*(T)}{m^*}&=&-\mathcal{A}\times \ln \left(\frac{E_F}{T}\right),\nonumber\\
\frac{\delta\tau_D^*(T)}{\tau_D^*}&=&\mathcal{A}\times \left[2\pi T\tau -
\ln\left(\frac{E_F}{T}\right)\right],\nonumber\\
\mathcal{A} &=& 
\left(1+\frac{15F_0^\sigma}{1+F_0^\sigma}\right)\frac{1}{4\pi^2\sigma_D},
\label{alpha-detailed}
\end{eqnarray}
for a system with two degenerate valleys.

The  equation for $\delta m^*(T)/m^*$ resembles the
one-loop renormalization of the effective mass (or $Z$)
in the RG theory
\cite{fink,CdCL98,fink02}. Our numerical
simulations  show that within the relevant  interval
$T=(0.03 - 0.8)$\,K and $r_s\lesssim 6$, the $\ln T $
terms in Eqs.~(\ref{alpha-detailed}) can be replaced with
a $T$-independent constant.
 By combining the LK
result with the interaction-induced corrections and replacing all
terms   $\propto \ln T$ by a constant within our limited $T$
range, we obtain the following linearized equation  in the
ballistic regime for the short-range scattering $(\tau_D \sim
\tau)$:
\begin{equation}
-\ln\left[A_1(T,B_\perp) \right] \frac{B_\perp} {2\pi^2 m^*} =
T+T_D\left(1-2\pi \mathcal{A} T\tau\right)=T+T_D\left(1-\frac{1}{2}\frac{\delta\sigma(T)}{\sigma_D}\right).
\label{reduced LK+GM}
\end{equation}
In this case, 
 the $T-$dependent correction to the Dingle temperature, $\frac{\delta T_D(T)}{T_D}$, is one half of
 the interaction correction to the conductivity \cite{ZNAballictic} $\frac{\delta\sigma(T)}{\sigma_D}$
 (this factor $\frac{1}{2}$ originates from the
difference between the interaction corrections to the momentum relaxation and quantum scattering
times \cite{QuantumAndTransportTime}).
We note that the empirical procedure used for finding
$m^*$ in  our earlier paper (Ref.\,\onlinecite{gm02}) was based on
the assumption that $T_D^*=T_D\left[ 1-
\delta\sigma(T)/\sigma_D\right]$, which differs from Eq.~(\ref{reduced LK+GM})
 by a factor of $\frac{1}{2}$.

At relatively high densities (which correspond to $r_s <4$), the
corrections to the LK result are insignificant within the studied
$T$-range. As $r_s$ increases, the temperature dependences of the
oscillation magnitude
predicted by the LK theory Eq.~(\ref{LK}) and the interaction
theory\cite{41} 
 start
deviating from each other. 
The values of
$|F_0^\sigma |$ extracted from SdH data using Eq.~(\ref{reduced LK+GM}) are
{\em larger } than those obtained with the LK theory but smaller than $|F_0^\sigma |$ obtained with
 the empirical procedure of Ref.~\onlinecite{gm02}  (e.g., at $r_s= 6.2$, the values  of $F_0^\sigma$
obtained according to Eq.~\ref{reduced LK+GM} and
the empirical procedure of Ref.~\onlinecite{gm02}
are 
$-0.40$ and $-0.45$,
respectively).
We have reanalyzed the data of Ref.\,\onlinecite{gm02} using
Eq.~(\ref{reduced LK+GM}) and compared the
corresponding values 
with $F_0^\sigma(n)$ extracted from the
$\sigma(T,B_\parallel)$ dependences using the ZNA theory 
(see below).

\subsubsection{Valley splitting and intervalley scattering}
The analysis of SdH oscillations
 using Eq.~(\ref{SdHbyLK}) also allowed us to estimate the
energy splitting $\Delta_V$  between the valleys. A non-zero
valley splitting  results in the beating
 of SdH oscillations  \cite{Valley_beats}.  Figure 1\,(a,b)
shows the SdH oscillations for samples Si6-14
and Si1-46 (the electron densities  are $6.1\times
10^{11}$cm$^{-2}$ and $1\times 10^{12}$cm$^{-2}$, respectively).
The amplitude of SdH oscillations normalized by the first harmonic
$A_1$ is expected to be field-independent if $\Delta_V=0$.  A
noticeable reduction of the SdH amplitude observed for both
samples at small fields can be attributed to a finite valley
splitting.  Although the node of SdH oscillations (expected at
$B\approx 0.15$\,T) cannot be resolved for samples with mobilities
$\mu \cong 2\,$m$^2$/Vs, $\Delta_V$ can be estimated from fitting of the $B-$dependence of the SdH
amplitude with Eq.\,(\ref{SdHbyLK}) modified for the case of a
finite $\Delta_V$: 
$\Delta_V\cong 0.4$\,K for sample Si6-14 and 0.7\,K for
Si1-46. This estimate  provides the upper
limit for $\Delta_V$  at $B=0$: in nonzero
 $B_\perp $ fields, $\Delta_V$  may be enhanced by the inter-level
interaction effects \cite{ando,42}.

The intervalley scattering rate for sample Si6-14 was measured
earlier in Ref.\,\onlinecite{43}  by analyzing the
weak-localization  (WL) magnetoresistance. It was found that $\tau_v$  
is temperature-independent and the ratio  $\tau_v/\tau$ decreases
monotonically with increasing electron density. For Si6-14 at $n =
(3-6)\times 10^{11}$\,cm$^{-2}$, $\tau_v \cong 20$\,ps
($\tau_v^{-1} \cong 0.36$\,K) is approximately ten times greater
than the transport time $\tau \cong   2$\,ps.

\subsubsection{Relaxation time $\tau$ and the band mass $m_b$}
\label{m_b} The momentum relaxation time $\tau$  was
determined  from the Drude conductivity 
$\sigma_D=n\tau/m_b$, which was
 found by extrapolating the quasi-linear $\sigma(T)$
dependence observed in the ballistic regime to $T = 0$. Note that
in order to extract $\tau$  from the Drude conductivity, one
should use the bare mass $m_b$: according to the Kohn theorem, the
response of a translationally-invariant system to the
electromagnetic field is described by $m_b$ in the presence of
electron-electron interactions; this result also holds for weak
disorder ($E_F\tau\gg1$). It is worth mentioning that several
prior publications \cite{30,29,28}, including our paper \cite{29},
incorrectly used $m^*$ instead of $m_b$ to estimate $\tau$  from
$\sigma_D$; this affects the value of the fitting parameters extracted from comparison with the theory
\cite{26} as shown below.

The textbook value \cite{ando} for the light electron mass in
 bulk Si is
$m_b^{(3D)}\approx 0.19m_e$. For inversion layers on
(001)~Si-surface, Kunze and Lautz \cite{tunneling85} have obtained
$ m_b^{(2D)}/m_e=(0.19 - 0.22) \pm 0.02$ from tunneling
measurements. Our  recent $m^*(n)$ data
 obtained from  the analysis of SdH  oscillations
\cite{gm02} 
 over a wide range of densities $r_s=1.4-8.5$,
  can be fitted with a polynomial $m^*(r_s) = 0.205m_e(1 + 0.035r_s  +
0.00025r_s^4)$.
  These $m^*/m_e$ data agree well with earlier 
 values of $m^*$  extracted from SdH 
 oscillations \cite{fowler66,fang&stiles68,smith&stiles72,35}
in narrower ranges of densities. By extrapolating the
polynomial $m^*(r_s)$ to $r_s=0$
 we obtain $m_b^{(2D)}/m_e=0.205\pm 0.005$, the value which
we adopted throughout the paper \cite{38}; available measurements
of the cyclotron resonance \cite{CR} do not contradict and do not
refine this value.

In principle, the aforementioned complete characterization of
samples allows us to compare the $\Delta\sigma(T,B)$ dependences
 with the theory \cite{26} without
any fitting parameters (with a caveat that the theory does not
take into account the intervalley scattering, see the discussion
below).  However,
throughout this paper we adopt an
equivalent, but more convenient procedure:
 for each electron density, $F_0^\sigma(n)$  will be considered
as a single parameter for fitting $\Delta\sigma(T,B)$, and these values of $F_0^\sigma$ will be
compared with the corresponding values obtained from the
SdH oscillations \cite{gm02}.

\subsection{Temperature dependences of the conductivity at $B_\parallel=0$}

The temperature dependences of the conductivity  $\sigma(T)$ for
sample Si6-14 are shown in Fig.~2. In these measurements, we
applied a fixed $B_\perp =0.1$\,T that is sufficient to suppress
the temperature dependence of the WL correction in the
studied temperature range.  The $\sigma(T)$ dependences are
non-monotonic for all studied densities [$n=(1.8-15)\times
10^{11}$cm$^{-2}$ for Si6-14 and $n=(10-15)\times
10^{11}$cm$^{-2}$  for Si1-46]: a quasi-linear increase of
$\sigma$ with cooling, observed
down to $\sim 0.5$\,K, is replaced at lower $T$ with a decrease of
$\sigma$. Note that in our previous experiments \cite{29}, we
observed   a trend of $\sigma(T)$ saturation
 at  $T <0.4$\,K
rather than the decrease of the conductivity.  One of the reasons
for  this might
 have been ``heating'' of electrons by high frequency noise:
only after thorough filtering of all 
 leads connected
to the sample 
were we able to decrease  the
 electron temperature down to $\sim 30$\,mK.
Similar downturn of $\sigma(T)$, although at much lower
temperatures, has been recently observed in high-$\mu$  GaAs FETs
at low electron densities \cite{huang07_pGaAs}. We
note also that earlier, a downturn of $\sigma(T)$   was observed
in Si-MOSFETs \cite{gmax} but at much higher electron densities
($n>30\times 10^{11}$cm$^{-2}$, $r_s <1.4$) and at much higher
temperatures $T\sim 10$K ($T\tau \approx 1$). For such high
densities the interaction corrections to $\sigma$  become negative
and the downturn of $\sigma(T)$ was related to the
 crossover between  a
``metallic''  high-temperature $\sigma(T)$ dependence (that is due
to electron--phonon and intersubband scattering effects)
 and resulting ``insulating'' low-temperature $\sigma(T)$ dependence
 (that is due to weak localization and negative interaction corrections
 contributions to $\sigma(T)$).

Below we use the following strategy for analyzing the
$\Delta\sigma(T)$ dependences. First, we find $F_0^\sigma$  by
fitting the quasi-linear
 $\sigma(T)$ dependences   observed in the
ballistic regime ($T > 0.5$\,K) with Eqs.~(\ref{1}-\ref{3}).
 The effect of valley splitting and intervalley scattering on
 $\Delta\sigma_{ee}(T)$ can be
 neglected at $T\gg \Delta_V, \tau_v^{-1}$ and the analysis
 is significantly simplified.
 The corresponding values of  $F_0^\sigma(n)$ are shown in Fig.~5.
 The $T$ range available for fitting in this regime ``shrinks'' rapidly at low $n$:
 the growth of $|F_0^\sigma |$  and decrease of $E_F$  with decreasing $n$,
lead to violation of the condition $T\ll (1+F_0^\sigma)^2E_F$
(e.g., at   $n=1.8\times 10^{11}$cm$^{-2}$
 this occurs at temperatures above 2\,K).
 This might be one of the reasons for the observed
deviation of the high-temperature $\sigma(T)$
from the linear-in-$(T/E_F)$ theory \cite{26}. Also, the higher-order
corrections might become significant at low $n$ when
$\Delta\sigma(T)/\sigma_D \sim 1$ (see Fig.~2).

After finding the $F_0^\sigma$  values (which are temperature-independent
in the studied temperature range), we proceed with the analysis of the low$-T$
part of the $\sigma(T)$ dependences, where the crossover from $d\sigma/dT <0$
to $d\sigma/dT >0$  was observed.  We note that the crossover
occurs when the
temperature becomes smaller than
two characteristic temperature scales -
$\Delta_V$ and $\tau_v^{-1}$  - which
are of the same order of magnitude for the studied structures.
We emphasize
that according to the ZNA theory\cite{26}, the ballistic-diffusive
crossover
 should not lead to
the change of the sign of $d\sigma/dT$.
In contrast,  the valley
splitting and  the intervalley scattering
 may result in the sign change for $d\sigma/dT$ because these processes
 reduce the number of triplet components
at $T<\Delta_V$ and $T< \tau_v^{-1}$.

The theory \cite{26} takes into account
a finite  $\Delta_V$ but not  $\tau_v^{-1}$.
The  solid red curves in Fig.~2 are calculated for
$\Delta_V=0.4$\,K  (the estimated value of $\Delta_V$
 for sample Si6-14) and  $\tau_v^{-1}=0$.  It is clear that the change in the
 number of triplet components from 15 ($T \gg \Delta_V$) to 7 ($T \ll \Delta_v$)
 \cite{45,30} (see also Eq.~\ref{5})
is not sufficient to explain the shape of the  $\sigma(T)$ downturn.
 The effect of strong intervalley scattering is illustrated in Fig.~2 by dashed
 green curves calculated with 3 triplet components (to model roughly the
 case of $T\ll \tau_v^{-1}$ when the valleys are completely intermixed).
 In  the absence of a detailed theory that would account
 for intervalley mixing, we attempted to fit the experimental data with an
 empirical crossover function for the number of triplet components
 $N_{\rm triplet}(x)=9+6[\exp(-0.3/x)-\exp(-30x)]$,
 where $x=T\tau_v$.
 This crossover function
 provides correct
 asymptotic limits for $N_{\rm triplet}$: 3 at $T\ll \tau_v^{-1}$
 and 15 at $T\gg \tau_v^{-1}$.
 Figure 2 shows that using this function,
 we can reasonably well describe the shape of experimental $\sigma(T)$ dependences
for all studied electron densities.

\subsection{Temperature dependences of the conductivity at non-zero $B_\parallel$}

Better understanding of different contributions to
$\Delta\sigma(T)$ can be achieved by measuring the conductivity in
strong in-plane magnetic fields $B_\parallel \gg T/g_b\mu_B$. The
evolution of experimental dependences  $\sigma(T)$ with
$B_\parallel$ is shown in Fig.~3 for two samples at different
electron densities.  The theoretical curves in Fig.~3 were
calculated using the $F_0^\sigma(n)$  values extracted from the
analysis of $\delta\sigma_{ee}(T,B_\parallel=0)$ (see Fig.~2).
 The transport time $\tau$  was calculated for each $B_\parallel$ value
 from the Drude conductivity
$\sigma_D(B_\parallel)$ which in turn was estimated by
extrapolating the quasi-linear part of the $\sigma(T,B_\parallel)$
dependence  to $T = 0$.
 The observed behavior is in line with our analysis of the
 $\sigma(T)$ dependences in Section III B.
 Indeed, the magnitude of the triplet contribution is expected to be reduced when the
 Zeeman energy becomes greater than $T$.  This effect is more pronounced within
 the range $\Delta_V$, $\tau_v^{-1} < T < g_b\mu_B B$,
 where a strong magnetic field reduces the number of
 triplet components from 15 to 7.
 For example,   at $n =1\times 10^{12}$cm$^{-2}$ (see Fig.~3\,d)
 the ``metallic'' behavior  disappears
 at $T<1$\,K
 and  $B_\parallel =3$\,T, which is in
 agreement with  the theory.
 At lower $T$, the number of triplet components
 is smaller than 15 even at  $B_\parallel =0$ due to valley splitting and
intervalley scattering, and the effect of $B_\parallel$
on $\Delta\sigma(T)$ is less prominent.

\subsection{Magnetoconductivity}
To test the theoretical predictions on the magnetoconductivity
(MC) induced by in-plane magnetic fields, we also measured the
$\sigma(B_\parallel)$ dependences at fixed $T$. Similar
measurements have been performed in the past (see, e.g., Refs.
\onlinecite{29,30,gao02_pGaAs}), but no detailed comparison with the theory was
carried out at that time.
The 
 MC for sample Si6-14 over the field range $-4.5 <B_\parallel <4.5$\,T
is shown for different densities and temperatures
in Fig.~4.  In these measurements, special care
was taken to reduce the magnetic field component perpendicular to the plane
of the structure: even a $1^\circ$ misalignment between the sample's plane and the
magnet axis (which results in $B_\perp  \sim 50$\,G at $B_\parallel=3$\,T)
may be sufficiently strong
for suppressing the WL corrections at low $T$.  To eliminate $B_\perp$,
we used the cross-magnetic-field set-up\cite{37}.  For each value of $B_\parallel$,
we measured the dependence $\sigma(B_\perp)$ by sweeping $B_\perp$
and recorded the minimum value of  $\sigma(B_\perp)$
which corresponded to the zero WL  magnetoconductance and, thus, $B_\perp =0$.
This method allowed us to compensate $B_\perp$  with an accuracy better than 10\,G.

The theoretical $\Delta\sigma(B_\parallel)$  dependences (see
Eqs.~(\ref{5},\ref{6})), plotted in Fig.~4 as solid  curves,
 describe the
observed  MC very well in not-too-strong magnetic fields $g_b\mu_B
B_\parallel <0.2E_F$. Again, as in the case of fitting the
$\Delta\sigma(T)$ dependences, the only adjustable parameter was
the $F_0^\sigma(n)$  value extracted for each density from fitting
the MC
at high temperatures ($\approx 0.7$\,K) where the effects of
valley splitting or intervalley scattering on
$\Delta\sigma_{ee}(T,B)$ can be neglected.  Note that all the
theoretical curves plotted in Fig.~4 for the same $n$ were
calculated for a fixed  $F_0^\sigma(n)$, i.e. neglecting possible
dependence $F_0^\sigma(B_\parallel)$. The detailed analysis of the
spin susceptibility  $\chi^* \propto g^*m^*$ in strong magnetic
field, presented in Ref.\,\onlinecite{47}, shows that
the product $g^*m^*$ decreases with an increase of $B_\parallel$
by as much as $\sim 20\%$.  
 Our estimate shows that by ignoring the $g^*(B)$
dependence, we might reduce the 
 value of $|F_0^\sigma|$
    by $\sim 10\%$ (see below), which is close to the accuracy of
    extraction of $F_0^\sigma$  from the data in strong magnetic
    fields.
As $B_\parallel$ grows and/or $n$
decreases, the data start deviating from the theoretical curves
(see Fig.~4\,d); this deviation can be attributed to the violation
of the condition $g_b\mu_B B \ll E_F$ required for applicability
of Eqs.~(\ref{5},\ref{6}).

\subsection{The  $F_0^\sigma(n)$ dependence}

The $F_0^\sigma$  values obtained from fitting the
$\Delta\sigma(T)$ and $\Delta\sigma(B_\parallel)$ dependences with
the theory \cite{26} are shown in Fig.~5.  For comparison, we have
also plotted
the 
$F_0^\sigma$ values obtained from the analysis of SdH
oscillations measured for sample Si6-14 using the theories \cite{LK,41}
(see Section III A). The  $F_0^\sigma$ values obtained from fitting
$\Delta\sigma(T)$ are in good agreement with the corresponding
values extracted from the analysis of SdH data.

At the same time, the 
$|F_0^\sigma|$ values
 obtained  from fitting the   $\Delta\sigma(B_\parallel)$ dependences
at $r_s>4$ are systematically smaller than the
corresponding
 values obtained from fitting  $\Delta\sigma(T)$ and SdH oscillations.
 This trend was earlier reported in Refs.\,\onlinecite{29,30}.
 There are at least two factors that can
reduce this discrepancy. 
 One of them, a potential decrease of $g^*$ in strong $B_\parallel$,
 was mentioned in Section III D.  The other factor is more subtle.
 In our analysis, we neglected the dependence of the 
 WL correction
 $\delta\sigma_{\rm WL}$ on $B_\parallel$.  However, our measurements
 show that  $\delta\sigma_{\rm WL}$ decreases with an increase of the
 in-plane magnetic field, which leads to a positive
 magnetoconductance.
 There are at least two potential reasons for this dependence:
 (a)~the Si-SiO$_2$ interface roughness transforms a uniform in-plane
 field into a random perpendicular field (see, e.g.,
 Refs.\,\onlinecite{49,50,51}
 and references therein), and (b)~a finite extent of electron wave
 functions in the direction perpendicular to the plane of  a quantum
 well  causes
  sub-band mixing by the magnetic field and disorder
  (see, e.g., Ref.\,\onlinecite{52}
 and references therein).

 Phenomenologically, both effects can be described in terms of  a
 decrease of the dephasing length $L_\varphi$  with $B_\parallel$.
 For example,  from the analysis of the WL magnetoresistance measured for different values of $B_\parallel$ for sample Si6-14 at $n = 1\times 10^{12}$cm$^{-2}$ and $T=0.3K$, we have extracted $L_{\varphi}(B_\parallel=0)=1.3\mu m$ and $L_{\varphi}(B_\parallel=3T)=0.8\mu m$. 
 Our estimates show that the positive magnetoconductance associated
 with the suppression of  $\delta\sigma_{\rm WL}$ by $B_\parallel$ can
 account for $\sim 10\%$ of the observed $B_\parallel$-induced
 magnetoconductance.  After taking the dependences $g^*(B_\parallel)$
 and  $\delta\sigma_{\rm WL}(B_\parallel)$ into account, the 
$|F_0^\sigma|$ values extracted from the MC should be
 increased  by $\sim 20\%$. The corresponding downshift of the  $F_0^\sigma$ values
 extracted from the $\Delta\sigma(B)$ data is shown by arrows
in Fig.~5\,a,
 it significantly reduces the
 discrepancy between the values of $F_0^\sigma$
 extracted from the $B_\parallel=0$ data and the data measured at $B_\parallel=1-5$\,T.

Finally, in order to compare our data with other
available  $\Delta\sigma(T)$ data  for high-mobility Si~MOSFETs,
we have 
 used Eqs.~(\ref{1}--\ref{6}) to estimate the $F_0^\sigma$ values
from the quasi-linear $\sigma(T,B=0)$ dependences measured by
Shashkin et al.\cite{28} and Vitkalov et al.\cite{30}. In this
analysis, we estimated $\tau$ from $\sigma(T\rightarrow 0)$ using
the band mass
 rather than $m^*$. As for the valley splitting,
 we have used $\Delta_V=0.6$\,K and $1\,$K for
 the analysis of data from Ref.\,\onlinecite{28} and \onlinecite{30},
respectively.
 (The value $\Delta_V\sim (1 - 2)$\,K used in
 Ref.\,\onlinecite{28}  seems to be too large, as
 it would lead to
the  appearance of beating of SdH oscillations
 in the field range studied in Ref.\,\onlinecite{krav_SdH00}).
We have also taken into account the WL correction neglected in
both Ref.\,\onlinecite{28} and \onlinecite{30}. Figure~5\,b shows
that the $F_0^\sigma(r_s)$ values  estimated for different
Si~MOSFETs using the ZNA theory are in good agreement with each
other.

The $F_0^\sigma$ values obtained on the basis of the
interaction correction theory and plotted in Fig.~5 may be
compared also with the values of $F_0^\sigma= -\gamma_2/(1+\gamma_2)$
 predicted by the RG
theory and measured experimentally in
Refs.\,\onlinecite{knyazev06,krav_nature_07}. 
Extrapolation of our
 $F_0^\sigma(r_s)$ data
to lower densities, provides the value $F_0^\sigma \approx -0.5$ at $n \approx 1\times 10^{11}$cm$^{-2}$ ($r_s \approx 8$), which is 
 {\em smaller} than the value $F_0^\sigma= - 0.31$
($\gamma_2 = 0.45$)  predicted by the one-loop RG theory for
the temperature $T^{\rm max}$ corresponding to the $\rho(T)$
maximum \cite{fink02,knyazev06,krav_nature_07} (e.g., $T^{\rm max}
\approx 3$\,K for $n=1.2 \times 10^{11}$cm$^{-2}$).

The experimental test\cite{knyazev06,krav_nature_07} of the RG theory
was conducted at temperatures higher than that in the experiments described in the present work.
Within the framework of the RG theory, the
interaction parameter $\gamma_2$  is expected to increases with decreasing $T$\cite{fink02,knyazev06,krav_nature_07} and, in principle, it  can
reach at $T<1K$ the value of  $\sim 1$ which corresponds to $F_0^\sigma = -0.5$. (Note that the    factor-of-two 
 increase of $\gamma_2$ (from $0.45$ to $1$) is beyond the range of the applicability of the one-loop RG theory.)
Another problem is that the spin susceptibility
$\chi^* \propto g^*m^*$
 obtained from the  SdH data is
almost $T$-independent \cite{47}, in contrast to the expected increase of $\gamma_2$
(and, hence, $|F_0^\sigma|$ and  $g^*$) with cooling.
This contradiction can be resolved if the $T$-dependence of  $g^*$
is exactly compensated by the opposite $T$-dependence of $m^*$, so that
$\chi^*\propto g^* m^*$ remains almost constant.
The reason for this compensation
 is not clear and requires both experimental and theoretical
 studies.

\section{CONCLUSION}

Our experiments show that the low-$T$ behavior of the conductivity
of high-mobility (001)~Si~MOSFETs is well described by the theory
of interaction effects in systems with short-range disorder
\cite{26}. Over a wide range of intermediate temperatures
($\Delta_V, \tau_v^{-1}, g_b\mu_BB <T \ll E_F$), the interaction
effects are strongly enhanced in Si~MOSFETs due to the presence of
two valleys in the electron spectrum.  This factor, in combination
with the interaction-driven renormalization of the Fermi-liquid
 parameter $F_0^\sigma$,
leads to an increase of $\sigma$  with decreasing $T$.  At lower temperatures ($T<\Delta_V, \tau_v^{-1}, g_b\mu_BB, E_F$),
the triplet contribution to $\Delta\sigma_{ee}(T)$ is
significantly reduced due to valley splitting and/or intervalley
scattering. As a result, the ``metallic'' behavior of $\sigma$  is
replaced
 with   a more  conventional, ``insulating''
 behavior. The
$F_0^\sigma$  values obtained from fitting the experimental data
with the theory \cite{26} agree well with the $F_0^\sigma$  data
obtained  from the analysis of SdH oscillations in these samples.
However, it remains unclear how to reconcile the
$F_0^\sigma$ values obtained at low $n$ from fitting the $\sigma(T)$ and SdH
data by using the interaction correction theory with the
corresponding values obtained within framework of the RG theory.

We emphasize that for the detailed analysis
 of the interaction-induced contributions to the conductivity, it is important
 to measure such parameters as the valley splitting and intervalley scattering
 rate in independent experiments.
Finally,
 for a quantitative description
 of the interaction effects to the conductivity
 $\Delta\sigma_{ee}$  at low temperatures,
both the interaction correction theory and RG theory should be
extended to the case of a finite intervalley scattering rate.

\section{ACKNOWLEDGMENTS}
The authors are thankful to E.~Abrahams, I.~Aleiner, I.~Burmistrov, I.~Gornyi,
G.~Kotliar,
 and A.~Mirlin for illuminating
discussions. The research at Rutgers University was partially
supported by the NSF grant ECE-0608842.  The research at Lebedev
Institute was supported by RFBR, Programs of the RAS, Russian
Ministry for Education and Science, and the Program ``Leading
Scientific Schools''.

\begin{figure}[h]
\includegraphics[width=260pt]{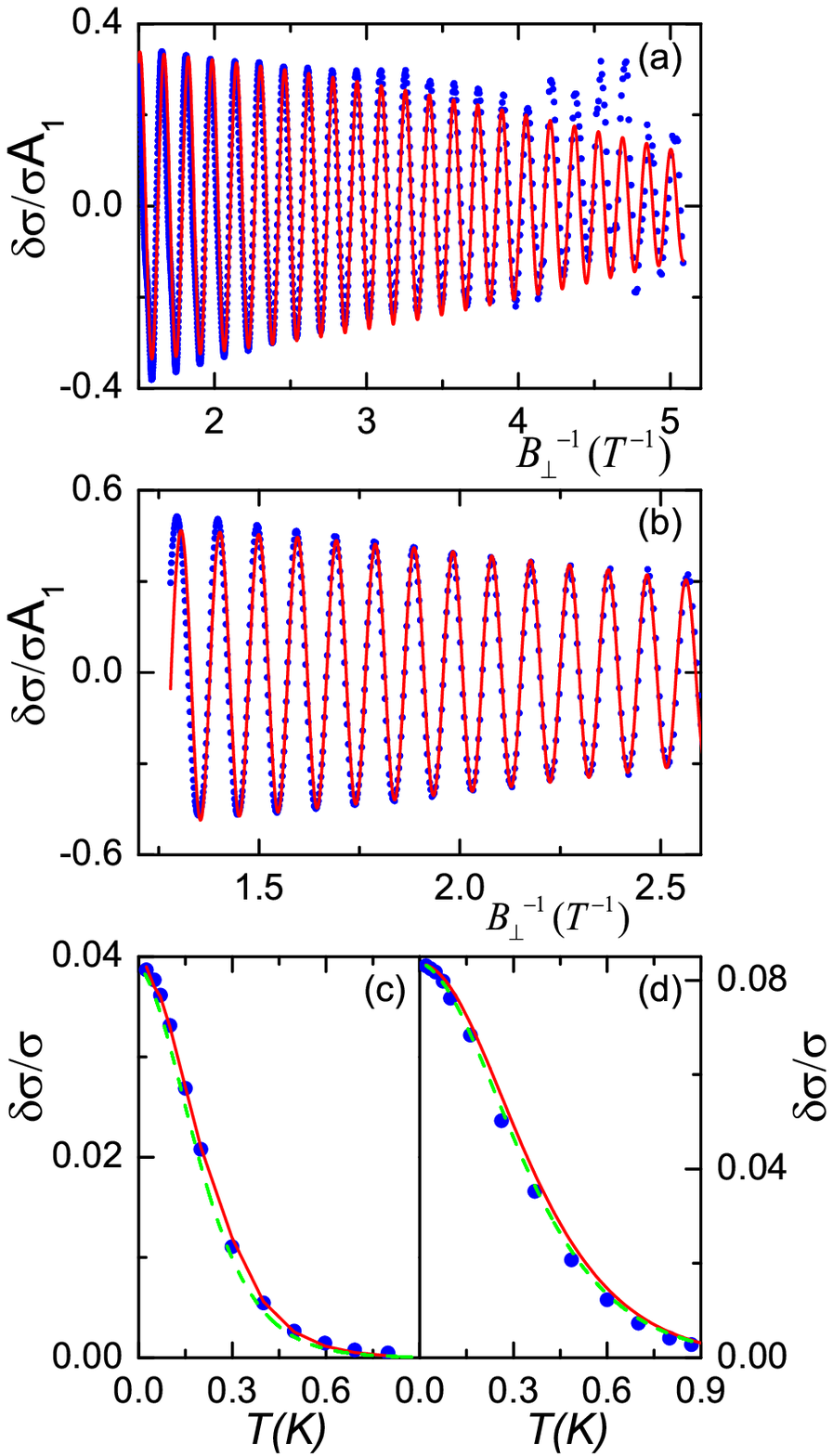}
\caption{(color online) Shubnikov - de Haas oscillations   normalized by the
amplitude of the first harmonic $A_1$: (a) sample Si6-14,
$n=6.1\times 10^{11}$cm$^{-2}$, $T=36$\,mK; (b)
sample Si1-46, $n=1\times 10^{12}$cm$^{-2}$,  $T=200$\,mK.
Dots represent the experimental data, solid
 curves - the theoretical dependences (7) modified for a finite $\Delta_V$
and calculated for  $\Delta_V=0.4$\,K and 0.7\,K  for samples
Si6-14 and Si1-46 respectively. Panels (c) and (d) show the
temperature dependences of the amplitude of SdH oscillations for
Si6-14 ($n=5.5\times 10^{11}$cm$^{-2}$) and Si1-46 ($n=1\times
10^{12}$cm$^{-2}$), solid  curves - the noninteracting LK-model
[Eqs.~(\ref{LK})], dashed curves - the fit based on the
interaction theory [Eqs.~(\ref{LK+GM}, \ref{alpha},
\ref{alpha-detailed})]. The linearized Eq.~(\ref{reduced LK+GM}) is indistinguishable from
Eqs.~(\ref{LK+GM}, \ref{alpha}, \ref{alpha-detailed})
within   the studied $T$ range and not
shown therefore.} \label{fig1}
\end{figure}

\begin{figure}[h]
\includegraphics[width=465pt]{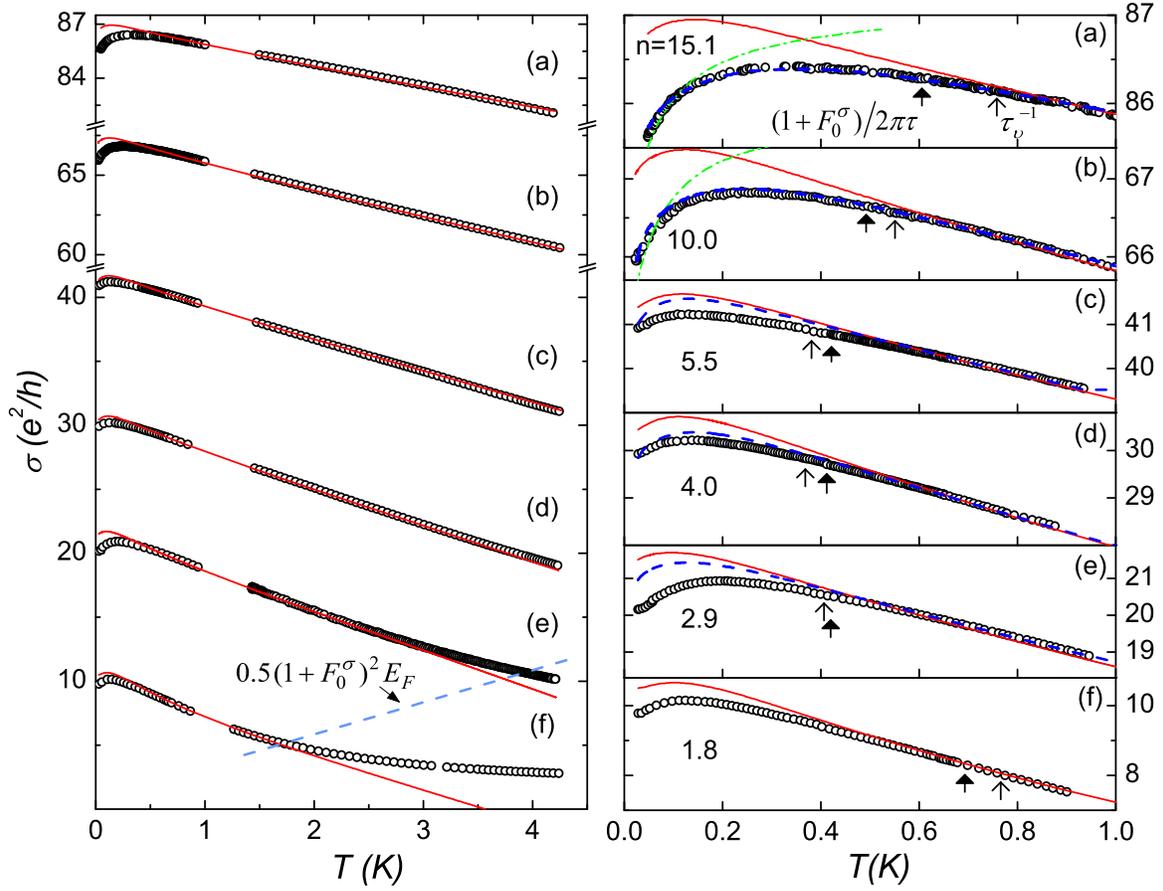}
\caption{(color online) Temperature dependences of the
conductivity $\sigma(T,B_\parallel=0)$ for sample Si6-14 at
various electron densities $n= 15.1, 10.0, 5.5, 4.0, 2.9, 1.8$, in
units of $10^{11}$cm$^{-2}$, from top to bottom. Circles show
experimental data, red
 curves - the theoretical dependences Eq.~(\ref{5})
calculated with  $\Delta_V=0.4$\,K. On the left panel, the dashed
blue curve corresponds to $T=0.5(1+F_0^\sigma)^2 E_F$, the applicability of
the ZNA theory is violated at a higher $T$. The right panel
shows the same data set within a narrower temperature interval,
the thin arrows correspond to  $T=\tau_v^{-1}$ and the thick
arrows - to the temperature of the crossover between ballistic and
diffusive regimes, $T^*=(1+F_0^\sigma)/2\pi \tau$.  The
dash-dotted green curves on
the right panel were calculated with three triplet components in
Eq.~(1) (the valleys are completely intermixed), the blue  dashed
 curves - $\sigma(T,B)=\sigma_D+\delta\sigma_C
+N_{\rm triplet}(T\tau_v)\times  \delta\sigma_T(T)$ with
$N_{\rm triplet}$  continuously varying between 3
 (for $T\ll \tau_v^{-1},\Delta_V$) and 15 (for $T\gg
\tau_v^{-1},\Delta_V$).}

\label{fig2}
\end{figure}

\begin{figure}[h]
\includegraphics[width=455pt]{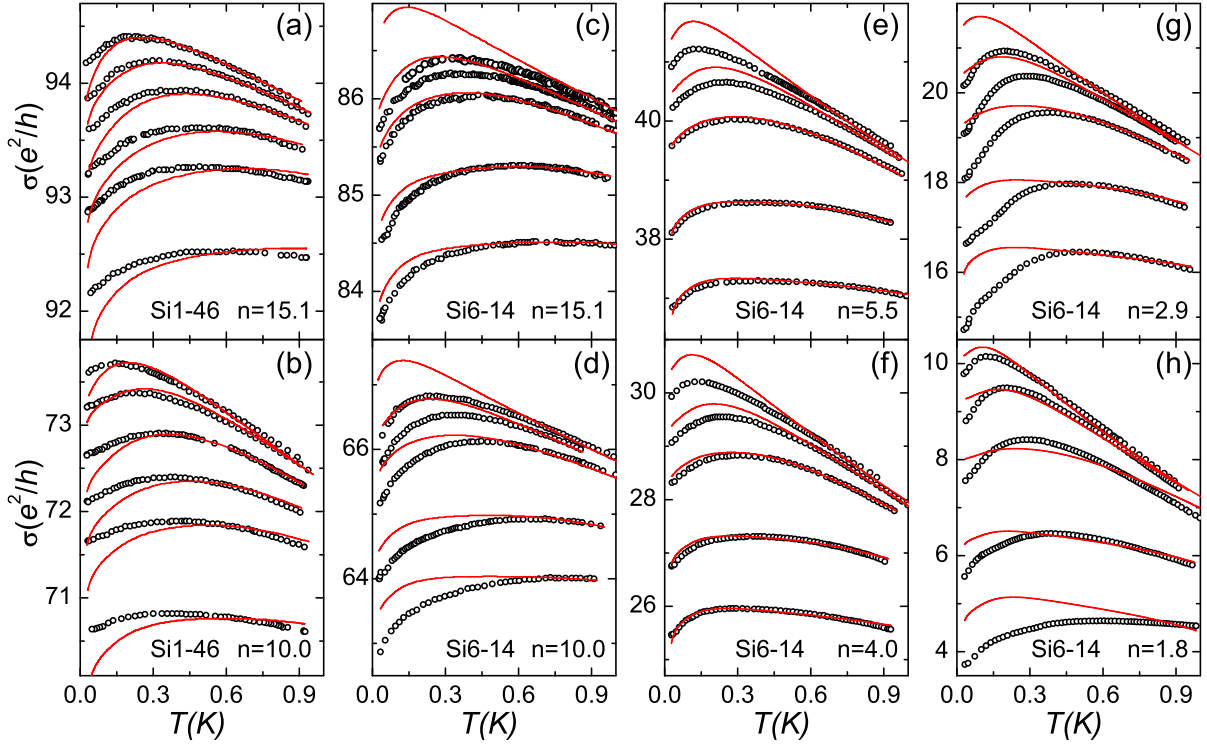}
\caption{(color online.) Temperature dependences of the conductivity for samples
Si1-46 [(a) and (b)] and Si6-14 [(c)$\div$(h)] in different in-plane
magnetic fields (from top to bottom, $B_\parallel=0, 0.6, 1, 1.5, 2, 3$\,T in panels (a) and (b), $B_\parallel=0, 0.6, 1, 2, 3$\,T in panels (c)$\div$(h)).
Experimental data are shown as circles, the solid  curves show the
theoretical dependences calculated for sample Si6-14 with
$\Delta_V=0.4$\,K  and for sample Si1-46 with $\Delta_V=0.7$\,K.
The $F_0^\sigma$  value is the only fitting parameter in
comparison with the theory \cite{26}, the corresponding values of
$F_0^\sigma$  are shown in Fig.~5. The values of $n$ are shown in
units of $10^{11}$cm$^{-2}$. } \label{fig3}
\end{figure}

\begin{figure}[h]
\includegraphics[width=207pt]{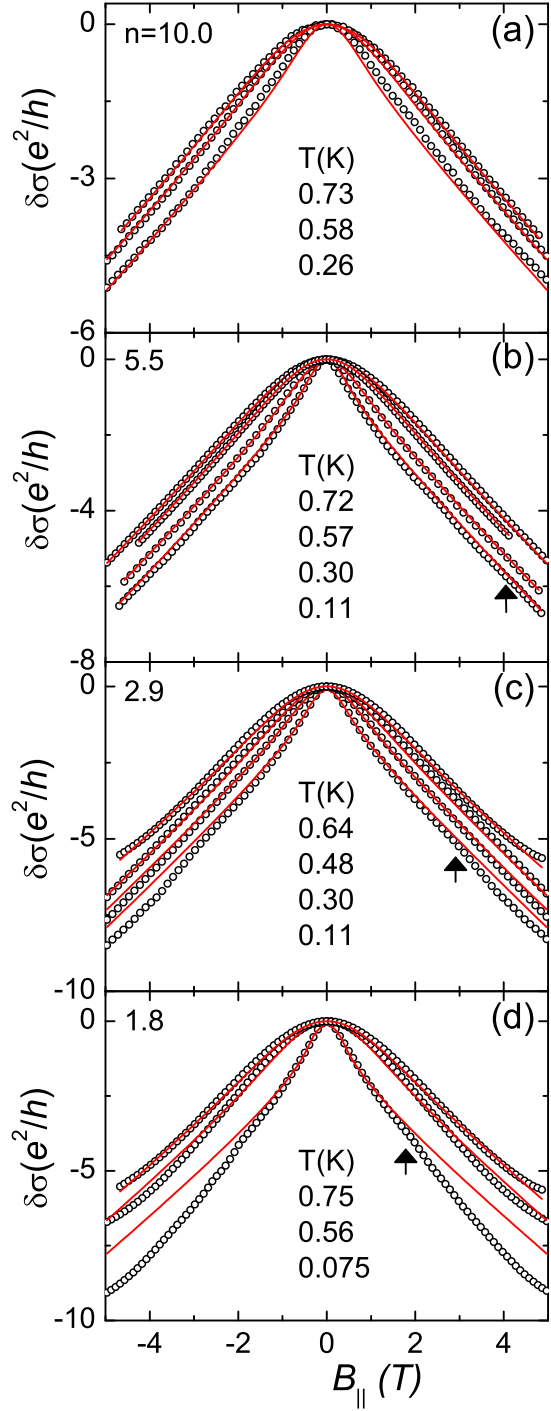}
\caption{(color online.) Magnetoconductance
for sample Si6-14 at different electron densities and
temperatures. Experimental data are shown by dots, the theoretical
dependences calculated according to Eqs.~(5,6) - by solid
 curves.
The $F_0^\sigma$  value is the only fitting parameter in comparison
with the theory \cite{26}, the corresponding values
of $F_0^\sigma$  are shown in Fig.~5.
Arrows indicate the fields corresponding to the condition
$g\mu_B B_\parallel/2E_F=0.1$.
The values of $n$ are shown in
units of $10^{11}$cm$^{-2}$.
}
\label{fig4}
\end{figure}

\begin{figure}[h]
\includegraphics[width=360pt]{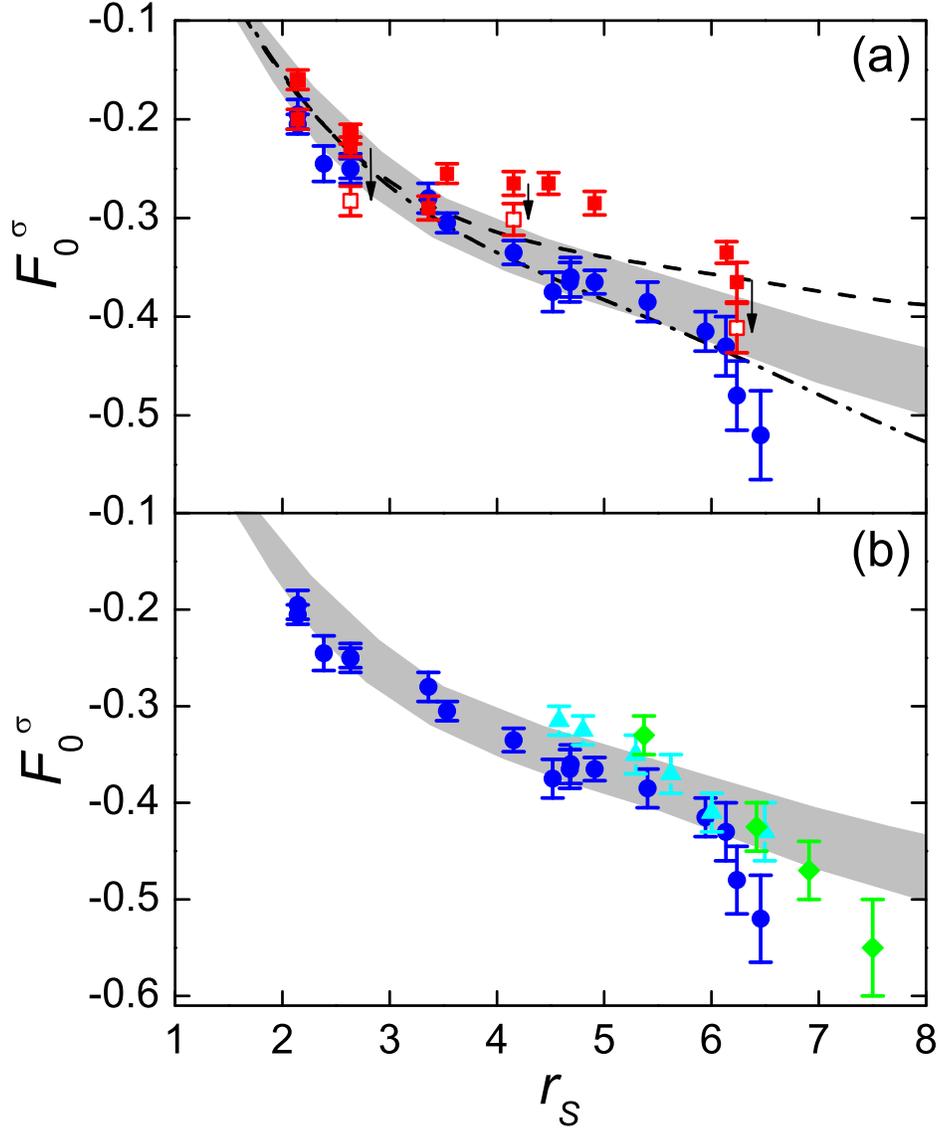}
\caption{(color online.) (a) The $F_0^\sigma$  values obtained from fitting the
$\sigma(T)$ and $\sigma(B_\parallel)$
dependences   with the theory \cite{26}
(blue and red
symbols, respectively).  Open red
squares show the shift of several $F_0^\sigma$ values
extracted from $\sigma(B_\parallel)$ if one takes into account
the $g(B_\parallel)$ dependence and suppression of the WL
corrections by $B_\parallel$ (see the text). 
The dashed curve corresponds to $F_0^\sigma(r_s)$ extracted from the SdH data\cite{gm02} using the LK theory, the dash-dotted curve - to the empirical approach used in Ref.~\onlinecite{gm02}. The shaded regions in panels (a) and (b) show the $F_0^\sigma(r_s)$
dependence (with the experimental uncertainty) obtained from fitting our
SdH data \cite{gm02} with the theory \cite{41}. (b) Comparison of  $F_0^\sigma$ values calculated from $\sigma(T,B=0)$ using the same fitting procedure (see the text): {\color{blue}\ding{108}} - present work, {\color{green}\ding{117}} and {\color{cyan}\ding{115}} - data recalculated from Refs.~\onlinecite{28} and \onlinecite{30}, respectively.}

\label{fig5}
\end{figure}


\begin{thebibliography}{99}
\bibitem{019me} Here we used a traditional value of $m_b$ in order to fasilitate the comparison with the results of other experimental works.
\bibitem{ZNAballictic}
Interaction correction to
the conductivity [Eqs.~(\ref{1}-\ref{3})] in the ballistic
regime has the form
$\delta\sigma(T)/\sigma_D=
T\tau[1+15F_0^\sigma/(1+F_0^\sigma)]/\pi\sigma_D$.
\bibitem{QuantumAndTransportTime}
The quantum correction to the transport scattering rate
$\tau^{-1}$ differs from that for the quantum scattering rate
$\tau_D^{-1}$ by the term $(1-cos(\phi))$ in the integrand, where
$\phi$ is the scattering angle. According to ZNA theory\cite{26},
the interaction corrections to the conductivity are determined by
backscattering events for which $\phi\approx\pi$, or
$(1-cos(\phi))\approx 2$.
\bibitem{38} Our revision of the analysis of SdH data
(described in the subsection~\ref{m*})
 modifies the $m^*$ values only
at low densities ($r_s>4$) and does not affect
$m^*(r_s\rightarrow 0)=0.205m_e$, adopted as the best estimate
for $m_b$.
\bibitem{45}  For a two-valley system with $\Delta_v\gg T$,
the total number of interaction channels is  $4\times 2=8$,
among them 1 singlet and 7 triplet terms.
\end{thebibliography}
\end{document}